\documentclass{article}
\usepackage[affil-it]{authblk}
\usepackage{graphicx}
\usepackage[space]{grffile}
\usepackage{latexsym}
\usepackage{textcomp}
\usepackage{longtable}
\usepackage{multirow,booktabs}
\usepackage{amsfonts,amsmath,amssymb}
\usepackage[numbers,sort&compress,square]{natbib}
\usepackage{url}
\usepackage{hyperref}
\hypersetup{colorlinks=false,pdfborder={0 0 0}}
\newif\iflatexml\latexmlfalse
\DeclareGraphicsExtensions{.pdf,.PDF,.png,.PNG,.jpg,.JPG,.jpeg,.JPEG}

\usepackage[utf8]{inputenc}
\usepackage[ngerman,english]{babel}
\usepackage[letterpaper, portrait, margin=1in]{geometry}
\usepackage{float}

\begin{document}

\title{Active mechanics reveal molecular-scale force kinetics in living oocytes}

\date{\today}
\author{}
\maketitle


\noindent
 \textbf{Wylie W. Ahmed}$^{1,2,7,\dagger}$, \textbf{\'{E}tienne Fodor}$^{3,7}$, \textbf{Maria Almonacid}$^{4,7}$, \textbf{Matthias Bussonnier}$^{2}$,  \textbf{Marie-H\'{e}l\`{e}ne Verlhac}$^4$, \textbf{Nir Gov}$^5$, \textbf{Paolo Visco}$^3$, \textbf{Fr\'{e}d\'{e}ric van Wijland}$^3$, \textbf{Timo Betz}$^{2,6}$
 
\noindent
$1$ Department of Physics, California State University, Fullerton, California 92831, USA \\
$2$ Laboratoire Physico-Chimie Curie, Institut Curie, PSL Research University, CNRS UMR168, 75005, Paris, France; Sorbonne Universit\'{e}s, UPMC Univ Paris 06, 75005, Paris, France \\
$3$ Laboratoire Mati\`{e}re et Syst\`{e}mes Complexes, UMR 7057, Universit\'{e} Paris Diderot, 75013 Paris, France \\
$4$ CIRB, Coll\`{g}e de France, and CNRS-UMR7241 and INSERM-U1050, \'{E}quipe Labellis\'{e}e Ligue Contre le Cancer, Paris F-75005, France \\
$5$ Department of Chemical Physics, Weizmann Institute of Science, 76100 Rehovot, Israel \\
$6$ Institute of Cell Biology, Center for Molecular Biology of Inflammation, Cells-in-Motion Cluster of Excellence, M\"{u}nster University, Von-Esmarch-Strasse 56, D-48149 M\"{u}nster, Germany\\
$7$ equally contributing authors \\
$\dagger$ corresponding author (wahmed@fullerton.edu)

\pagebreak{}

\begin{abstract}
Active diffusion of intracellular components is emerging as an important process in cell biology. This process is mediated by complex assemblies of molecular motors and cytoskeletal filaments that drive force generation in the cytoplasm and facilitate enhanced motion.  The kinetics of molecular motors have been precisely characterized \textit{in-vitro} by single molecule approaches, however, their \textit{in-vivo} behavior remains elusive. Here, we study the active diffusion of vesicles in mouse oocytes, where this process plays a key role in nuclear positioning during development, and combine an experimental and theoretical framework to extract molecular-scale force kinetics  (force, power-stroke, and velocity) of the \textit{in-vivo} active process. Assuming a single dominant process, we find that the nonequilibrium activity induces rapid kicks of duration $\tau \sim$ 300 $\mu$s resulting in an average force of $F \sim$ 0.4 pN on vesicles in \textit{in-vivo} oocytes, remarkably similar to the kinetics of \textit{in-vitro} myosin-V.  Our results reveal that measuring \textit{in-vivo} active fluctuations allows extraction of the molecular-scale activity in agreement with single-molecule studies and demonstrates a mesoscopic framework to access force kinetics.
\end{abstract}%


\section{Introduction}
Living cells utilize motor proteins to actively generate force at the molecular scale to drive motion and organization in the crowded intracellular environment \cite{Howard_2009,24382887,17461730,20110992}. For example, force generation is critical to facilitate the basic tasks of living cells such as spatial organization, motility, division, and organelle transport.  Recently, ``active diffusion" has emerged as an important process in cell biology \cite{25774831,19699642,27003292,24361104,27003290}.  Active diffusion is the random motion of intracellular components that is driven by active metabolically powered forces (and not by thermal fluctuations). An interesting example occurs during early vertebrate development, namely oocyte meiosis where active diffusion drives nucleus centering \cite{25774831}.  

\begin{figure}[p]
\begin{center}
\includegraphics[width=1\columnwidth]{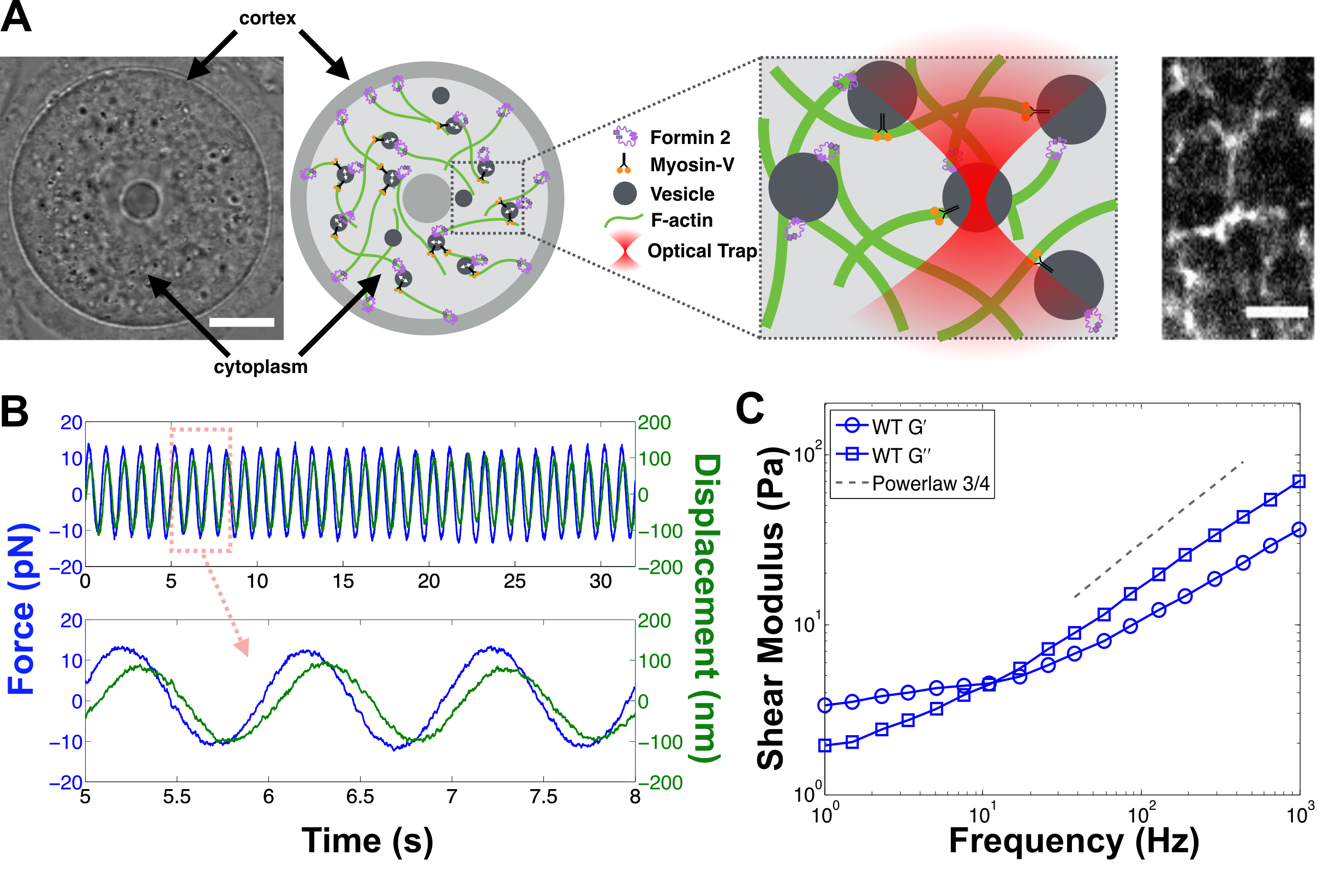}
\caption{{\label{fig:fig1}
\textbf{Intracellular mechanics surrounding endogenous vesicles
in living oocytes.} (A) Mouse oocytes are large spherical cells that
have a well separated cortex and cytoplasmic-skeleton composed of biopolymer filaments (brightfield image, left)(scalebar = 20 $\mu$m). During prophase-I it
contains a dynamic actin-myosin-V meshwork that drives vesicle motion (schematic shown in center and fluorescent image of actin filaments on right)(scalebar = 5 $\mu$m). Endogenous
vesicles embedded in the cytoplasmic-skeleton are trapped using optical tweezers (zoomed inset). (B) Once a vesicle is trapped, the mechanical properties
of the local environment can be measured by active microrheology (AMR)
where a sinusoidal oscillating force is applied to the vesicle (blue)
and the resulting displacement of the vesicle is measured (green).
The viscoelastic shear modulus ($G^{*}$) is calculated from this
force-displacement measurement via the Generalized Stokes-Einstein relation. (C)
The mechanical properties surrounding vesicles in the cytoplasmic-skeleton
of oocytes exhibits power-law behavior with frequency scaling $G^{*}\propto f^{0.75}$. This shows that the cytoplasmic-skeleton in oocytes
can be modeled as a semi-flexible polymer network.%
}}
\end{center}
\end{figure}

Oocytes are immature female gametes that are destined to be fertilized and grow into a fully functioning organism.  They are large cells (80 $\mu$m in diameter) of spherical geometry with a thick actin-rich cortex and a well separated cytoplasmic-skeleton (Fig. \ref{fig:fig1}A). The cytoplasmic-skeleton of a mouse oocyte is a composite material that includes actin filaments, microtubules, and intermediate filaments.  The actin network is composed of long unbranched filaments polymerized from the surface of vesicles that harbor actin nucleating factors \cite{21983562}.  These vesicles act as nodes to create an inter-connected actin network that is uniform in density throughout the cytoplasm and un-polarized \cite{23873150}; Microtubules are organized in small seeds and do not form long filaments during prophase-I \cite{23954884}; And while it is probable that intermediate filaments are present \cite{nikolova2012}, little is known about their structure or function in mouse oocytes.

Correct nuclear positioning during prophase-I requires precise spatial and temporal coordination of the cytoskeleton and molecular motors \cite{23851486,23873150}. Recent work has shown the importance of mechanical processes in oocyte development \cite{23851486,23873150,25774831}.  Proper spindle positioning requires the oocyte cortex to soften and exhibit plastic deformation \cite{23851486}, and motion of the cytoplasmic actin network is closely regulated by myosin-V motors \cite{23873150}.  Motion of large objects, such as positioning of the spindle \cite{23851486} or nucleus \cite{25774831}, must be driven by a physical force generated by nonequilibrium processes. For instance, we have recently shown that in prophase-I myosin-V driven vesicle motion maintains a soft cytoskeleton and generates a force to center the nucleus \cite{25774831}. Here we focus on the physical understanding of the active forces that generate vesicle motion, and how the detailed quantification and modeling of this activity can successfully predict characteristic parameters of the active forces, such as timescales, velocities and effective forces.   

To investigate the nonequilibrium mechanical activity in living cells, it is necessary to independently measure the active force generation and the local mechanical properties to understand how objects move within the complex intracellular environment \cite{11724945,14611619,17234946,25126787,Gallet_2009,15937489,18764230,Turlier_2016,Campas_2016,Sugimura_2016}. This approach was introduced to study active processes in hair bundles \cite{11724945} and stress fluctuations in cells \cite{14611619}. Subsequent studies utilized this concept to investigate cross-linked actin-myosin-II networks in reconstituted systems \cite{17234946}, beads embedded in living cells \cite{25126787,Gallet_2009,15937489,18764230}, and red blood cell membranes \cite{Turlier_2016}. Alternative methods of estimating forces \textit{in-vivo} include FRET force sensors, droplet-based sensors, and force inference as recently reviewed \cite{Campas_2016,Sugimura_2016}.

In this study, we measure the actively driven motion of endogenous vesicles in the largely unexplored cytoplasmic-skeleton of \textit{in-vivo} mouse oocytes and demonstrate a framework to quantify the underlying driving forces. Endogenous vesicles embedded in the oocyte cytoplasm experience local elastic caging. The vesicles undergo actively driven motion via a sparse actin-myosin-V force generating network that remodels the environment. The actin-myosin-V network is the key element that generates force for nuclear positioning \cite{25774831}, yet the underlying physical processes related to active force generation are not well understood.

To investigate cytoplasmic force generation, we combine optical tweezer based active microrheology (AMR) and laser-tracking interferometry with our theoretical framework to quantify the \textit{in-vivo} active mechanical processes and relate them to the underlying stochastic force kinetics (amplitude and temporal correlations). To study the activity in mouse oocytes, we examine the cytoplasmic-skeleton during prophase-I, when the actin-myosin-V network is the dominant source of activity \cite{25774831}. By using statistical mesoscopic measurements on endogenous vesicles, and theoretical modeling, we extract the molecular-scale force kinetics (average power-stroke duration, displacement, force, and velocity) of \textit{in-vivo} myosin-V.

\section{Materials and Methods}

\subsection{Fmn-/- and Myosin-V dominant negative mouse oocytes}
To study the effect of the actin-myosin-V network we utilize two conditions to independently perturb the actin cytoskeleton (Fmn-/-) and the myosin-V activity (MyoV(-)). Fmn-/- mouse oocytes have no detectable cytoplasmic actin filaments, as confirmed by several independent studies \cite{18848445,19062278,21620703,23851486}. The reduced cytoplasmic actin has also been confirmed via cytochalasin-D treated mouse oocytes, which exhibit the same phenotype and mechanical properties as Fmn-/- \cite{16989804} (Fig. S1). For myosin-V dominant negative experiments (MyoV(-)), oocytes were injected with cRNAs using an Eppendorf Femtoject microinjector as published previously \cite{25774831}. Oocytes were kept in prophase-I arrest for 1-3 hours to allow expression of fusion proteins. The myosin-Vb dominant negative construct corresponds to a portion of the coiled-coil region of the myosin-Vb that mediates dimerization of the motor \cite{21983562}.  We believe this construct binds to the coiled-coil region of endogenous myosin-Vb resulting in impaired motor dimerization.  Thus, myosin-Vb is still able to bind to vesicles and actin filaments but does not actively generate force. It is specific to myosin-Vb and has been shown to stop vesicle motion in mouse oocytes as efficiently as brefeldin A (BFA), which is a general traffic inhibitor \cite{25774831}.  Note that MyoV(-) oocytes were referred to as WT + MyoVb tail oocytes in a previous study \cite{25774831}.

\subsection{Oocyte Preparation}
Oocytes were collected from 11 week old mice OF1, 13 week old C57BL6
(WT) or 15 week old Fmn-/- female mice as previously described \cite{7600950}
and maintained in prophase-I arrest in M2+BSA medium supplemented
with 1 $\mu$M Milrinone \cite{16715549}. Live
oocytes were embedded in a collagen gel to stop movement of the overall
cell during measurements. Collagen gel was made by mixing M2 medium
(33.5$\mu$L), 5X PBS (10 $\mu$L), NaOH (1M, 0.9 $\mu$L), collagen
(3.6 mg/mL, 55.6 $\mu$L) to obtain 100 $\mu$L of the final collagen
solution at 2 mg/mL with a pH$\sim7.4$. 20 $\mu$L of the collagen
solution was deposited on a coverslip and live oocytes were added.
The droplet was covered with another glass coverslip using Dow Corning
vacuum grease to minimize evaporation. The collagen
gel containing oocytes was polymerized in a humid environment at 37$^{\circ}$
C for 30 minutes.


\subsection{Optical Tweezer Setup}
The optical tweezer system utilizes a near infrared fiber laser ($\lambda=1064$
nm, YLM-1-1064-LP, IPG, Germany) that passes through a pair of acousto-optical
modulators (AA-Optoelectronics, France) to control the intensity and
deflection of the trapping beam. The laser is coupled into the beam
path via dichroic mirrors (ThorLabs) and focused into the object plane
by a water immersion objective (60x, 1.2 NA, Olympus). The condenser
is replaced by a long distance water immersion objective (40x , 0.9
NA, Olympus) to collect the light and imaged by a 1:4 telescope on
a InGaAs quadrant photodiode (QPD) (G6849, Hamamatsu). The resulting
signal is amplified by a custom built amplifier system (Oeffner Electronics,
Germany) and digitized at a 500 kHz sampling rate and 16 bit using
an analog input card (PCIe-6353, National Instruments, Austin, TX, USA). All control of the experimental
hardware is executed using LabVIEW (National Instruments). Optical
trapping of endogenous vesicles was calibrated similarly as in \cite{23820071,23404705},
where the high-frequency fluctuations ($f>500$ Hz) from the fluctuation and response measurements are aligned to determine the trap stiffness \cite{25126787,24876498}.  This approach assumes that the high frequency motion is dominated by thermal fluctuations so that the fluctuation dissipation theorem (FDT) holds in this regime. For direct measurement of
violation of the FDT, laser tracking
interferometry is used first to measure the spontaneous fluctuations
of the vesicle, followed immediately by active microrheology to measure
the mechanical properties of the local environment surrounding the
vesicle.

\subsection{Laser Tracking Interferometry}
The position of the endogenous particle is measured by back focal plane interferometry \cite{18084394}. It should be noted that deformable objects (e.g., giant unilamellar vesicles $10-100 \mu m$) undergo shape fluctuations that will manifest in the voltage measured by the QPD \cite{Betz_2012}. For small endogenous particles ($\sim1 \mu m$), it has been confirmed that shape fluctuations are small and laser interferometry can be used to track their position with nanometer precision.  This has been verified in mammalian cells \cite{10733956} and yeast \cite{Toli_N_rrelykke_2004}.  A recent study has also confirmed that laser tracking interferometry and the active-power spectrum calibration \cite{23820071} allow quantitative force-displacement measurements on vesicles\textit{ in-vivo} \cite{25229154}.  To confirm the validity of this assumption in mouse oocytes we investigated the deformability of endogenous vesicles and conclude that the majority of measured fluctuations come from active vesicle displacements (see Fig. S5-10). Additionally, since the endogenous vesicles serve as nodes integrated in the cytoplasmic meshwork, they are accurate reporters of the network mechanics and fluctuations \cite{15189896}.

\subsection{Data Analysis}

In the active microrheology (AMR) experiments we apply a known force,
$F_\text{ext}$, to an endogenous vesicle and measure the resulting displacement,
$u$. Using linear response theory, these are related to the material
response function $\chi (t)$ as $u(t)=\int_{-\infty}^{t}\chi(t-t^{\prime})F_\text{ext}(t^{\prime})dt^{\prime}$.
In Fourier space, we can directly calculate the response function as
$\tilde{\chi}(\omega)=\tilde{u}(\omega)/\tilde{F}_\text{ext}(\omega)$. To translate
the response function into a complex shear modulus $G^{*}$, we use the Generalized
Stokes-Einstein relation $G^{*}=1/[6\pi R\tilde{\chi}(\omega)]$, where $R$ is the average radius of the spherical tracers. It can be split into a real part $G'$ and an imaginary part $G''$, which respectively provide information about the elastic-like and viscous-like properties of the medium. For AMR measurements the laser power exiting the objective
was $\sim120$ mW. For the spontaneous fluctuations we measure the
motion, $u(t)$, of endogenous vesicles via laser interferometry (without
trapping the vesicle, laser power $\sim$ 1 mW) and calculate the power spectral density (PSD),
$\tilde{C}(\omega)=\int\langle u(t)u(0)\rangle\exp(i\omega t)dt$, as described
previously \cite{Betz_2012}. Briefly, we calculate the PSD by using
MATLAB (The Mathworks, USA) to take the Fast Fourier Transform (FFT)
of the vesicle position, $\tilde{u}=\mbox{FFT}(u)$. Then the PSD
is calculated as, $PSD=\frac{\tilde{u}\times\tilde{u}^{*}}{p\times s}$,
where $^{*}$ denotes complex conjugate, $p$ is the number of measurement
points, and $s$ is the sampling frequency. In systems at thermal
equilibrium the response function can be calculated from the PSD via
the FDT as is done for passive microrheology (PMR) \cite{Kubo_1966}.
Due to the particle size and range of the QPD measurement
our PMR measurements are restricted to a particle displacement of approximately 500 nm during the measurement interval.  If the particle left this regime, that part of the data was not analyzed.  The limitation to the linear regime of the detector does not imply that we measure only the linear regime of the force displacement relation.   AMR and PMR data analysis is carried out as done previously \cite{Turlier_2016}.


\section{Results}

    \subsection{Intracellular mechanics of the oocyte cytoplasmic-skeleton}
    To measure the local mechanical properties, we optically trap an endogenous vesicle embedded in the cytoplasmic-skeleton, apply an oscillatory force while measuring its displacement (Fig \ref{fig:fig1}B), and calculate the shear modulus via the Generalized Stokes-Einstein relation \cite{Mizuno_2008}.  Since vesicles serve as integrated nodes in the cytoplasmic-skeleton, they accurately reflect the mechanics and fluctuations of the network \cite{15189896} (details on vesicle tracking in SM). To calibrate the force on the vesicles we exploit the established observation that high-frequency fluctuations are thermal in origin \cite{24876498,25126787,23820071}. AMR directly measures the mechanical response function ($\tilde{\chi}$), which allows to determine the complex modulus ($G^{*}$) characterizing the viscous and elastic resistance of the composite cytoplasmic-skeleton including any contribution from actin, intermediate filaments, microtubules, and other structures present. We find that the cytoplasmic-skeleton of oocytes has strongly viscoelastic properties similar to semiflexible biopolymer networks \cite{10059905,Morse_1998}, exhibiting high-frequency power-law behavior ($G^{*}\propto f^{\alpha}$, where $\alpha \sim 0.75$) (Fig \ref{fig:fig1}C). Thus, the cytoplasmic-skeleton in oocytes is strongly frequency dependent, which must be accounted for when modeling its mechanical behavior. Extending the theoretical model into the viscoelastic regime with a strong frequency dependence is a necessary advancement to describe the intracellular mechanics. 
    
\begin{figure}[p]
\begin{center}
\includegraphics[width=0.5\columnwidth]{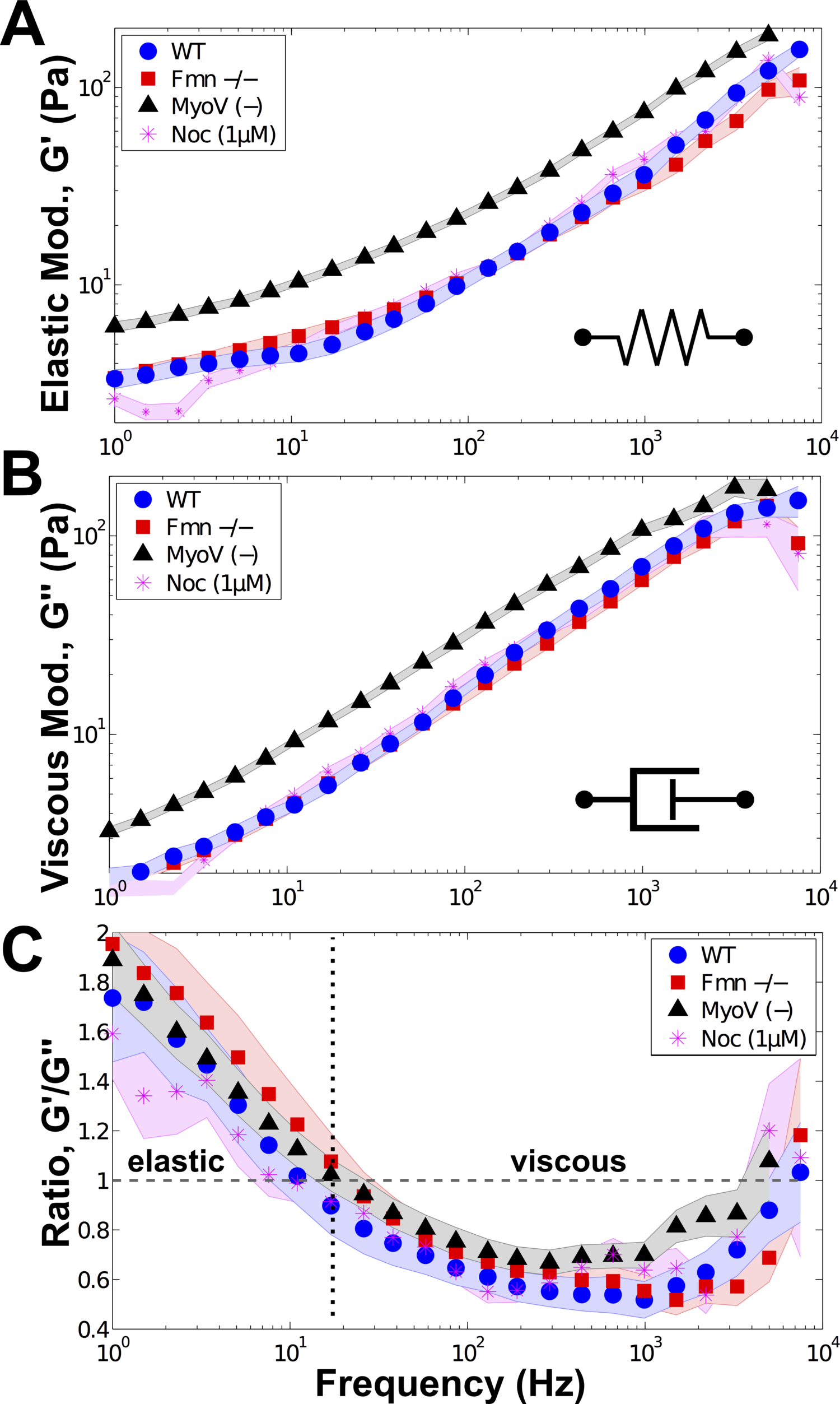}
\caption{{\label{fig:fig2}
\textbf{The cytoplasmic-skeleton of mouse oocytes is viscoelastic.} (A, B) Insets indicate that $G^{\prime}$
quantifies the elasticity and $G^{\prime\prime}$ quantifies the viscous
dissipation of the cytoplasmic-skeleton. The local mechanical properties
(elastic and viscous) surrounding vesicles does not change from WT (blue, $\circ$) when actin (red $\square$) or microtubules (magenta *) are absent. However, when myosin-V
is inactivated (gray $\triangle$), the cytoplasmic-skeleton stiffens significantly
(Kolmolgorov-Smirnov test, $p<1\times10^{-7}$), showing that the absence of an
active actin-myosin-V meshwork leads to a stiffer mechanical environment surrounding vesicles in
oocytes. (C) The
ratio of the elastic and viscous moduli ($G^{\prime}/G^{\prime\prime})$
in all oocytes shows that the cytoplasmic-skeleton is more elastic
at lower frequencies and more viscous at higher frequencies with a
crossover around $\sim 10-20$ Hz. This shows the highly viscoelastic nature
of the oocyte cytoplasmic-skeleton. (sample size = WT: 11 cells, 32 vesicles; Fmn-/-: 10 cells,
33 vesicles; MyoV(-): 23 cells, 69 vesicles; Noc$(1\mu$M): 8 cells, 52 vesicles; shaded region indicates
SEM) (note: data at 10 Hz in panel A and B are presented in a different
context in \cite{25774831})%
}}
\end{center}
\end{figure}

    To dissect the mechanical contributions of cytoskeletal filaments in oocytes we first perform AMR on formin-2 knockout oocytes (Fmn-/-), which removes cytoplasmic actin filaments nucleated by formin-2 \cite{18848445}, as confirmed via cytochalasin-D experiments as an alternative method of removing actin (Fig. S1). We find that the mechanical properties do not change compared to WT, indicating that the actin network does not provide significant mechanical resistance in the oocyte (Fig. \ref{fig:fig2}A, B). This behavior is presumably because the actin network in WT oocytes is sparse (Fig \ref{fig:fig1}A, right and Fig. S2) with a mesh size of $5.7 \pm 1.9 \mu$m \cite{19062278}. Depolymerizing the microtubules by nocodazole treatment (1 $\mu$M) does not significantly affect the mechanical properties (Fig. \ref{fig:fig2}A, B),  indicating that microtubules also do not provide significant mechanical resistance in the cytoplasmic-skeleton. This result is expected since prophase-I microtubules form small seeds instead of long force-bearing tubule structures \cite{23954884}. These combined results suggest that there are other mechanical scaffolds contributing to the stiffness in the cytoplasm of mouse oocytes.  
    
    While our results show that actin does not contribute significantly to the mechanical resistance, it is known that the actin-myosin-V meshwork activity plays a critical role for self-organization such as positioning of the meiotic spindle \cite{18848445,19062278} and nucleus centering in mouse oocytes \cite{25774831}. To independently probe the activity of this network we impaired the myosin-V motor force generation by de-activating it via microinjection of a dominant negative construct (MyoV(-))\cite{21983562,25774831} (see Materials and Methods). Under these conditions the mechanical properties of the cytoplasmic-skeleton stiffens significantly as evidenced by the shift upwards in the elastic and viscous moduli (Fig \ref{fig:fig2}A, B). This increased mechanical stiffness correlates with a higher density actin meshwork that is observed in the absence of myosin-V activity (Fig. S2 in SM) and can be explained by increased cross-linking \cite{23873150}.  Actin networks are known to stiffen significantly upon addition of crosslinks, as long as the inter-crosslink distance is smaller than the persistence length. 
    
The overall behavior of the cytosplasmic-skeleton in oocytes is strongly viscoelastic where neither the elastic or viscous components dominate by a large margin.  At the lowest probed frequencies the behavior is slightly more elastic with a crossover ($\sim 10-20$ Hz) to more viscous at higher frequencies.  However both components of the shear moduli are always within a factor of two at the observed frequencies (Fig. \ref{fig:fig2}C).  This is in contrast to adherent cells which typically show an overall weaker frequency dependence and are dominantly elastic \cite{24094397,25126787}.

    \subsection{Quantifying \textit{in-vivo} active force fluctuations}
To quantify nonequilibrium activity we use laser-tracking to precisely
measure the spontaneous motion of vesicles in the cytoplasmic-skeleton
with high spatio-temporal resolution \cite{18084394,10733956,Toli_N_rrelykke_2004} (see Materials and Methods). Both the spontaneous motion and the local mechanics are measured for each individual vesicle via active microrheology and laser interferometry.  This allows direct comparison between the local mechanical environment experienced by the vesicle and the active force that is driving its motion since both measurements are made \textit{in-situ}.  This is critical in order to extract information about the molecular-scale processes. In the absence of biochemical activity the motion of the vesicles would be due
to purely thermal agitation, which is fully determined by the mechanical
properties of the system. This basic relation is given by a fundamental
theorem of statistical mechanics known as the fluctuation-dissipation
theorem (FDT) \cite{Kubo_1966}. The FDT relates the small fluctuating
motion of the vesicles to the mechanical properties of the surrounding
environment by, $\tilde{\chi}^{\prime\prime}=\pi f\tilde{C}/k_{B}T$,
where $\tilde{\chi}^{\prime\prime}$ is the dissipative part of the
mechanical response, $\tilde{C}$ is the power spectral density of
vesicle motion, $f$ is frequency,  $k_{B}$ is the Boltzmann constant, and $T$ is the temperature (see Table 1 for list of symbol definitions).
However, in living cells the presence of biochemical activity gives
rise to active forces (e.g. a nonequilibrium process) driven by energy consuming
processes \cite{11724945,14611619,26025677}. In other words, the force driving the motion of particles
in a living cell (e.g. the oocyte) has two contributions:
(1) a passive (purely thermal) contribution described
by classical equilibrium physics; (2) an active contribution that is biochemically regulated and cannot be understood
via equilibrium physics.

\begin{figure}[p]
\begin{center}
\includegraphics[width=0.8\columnwidth]{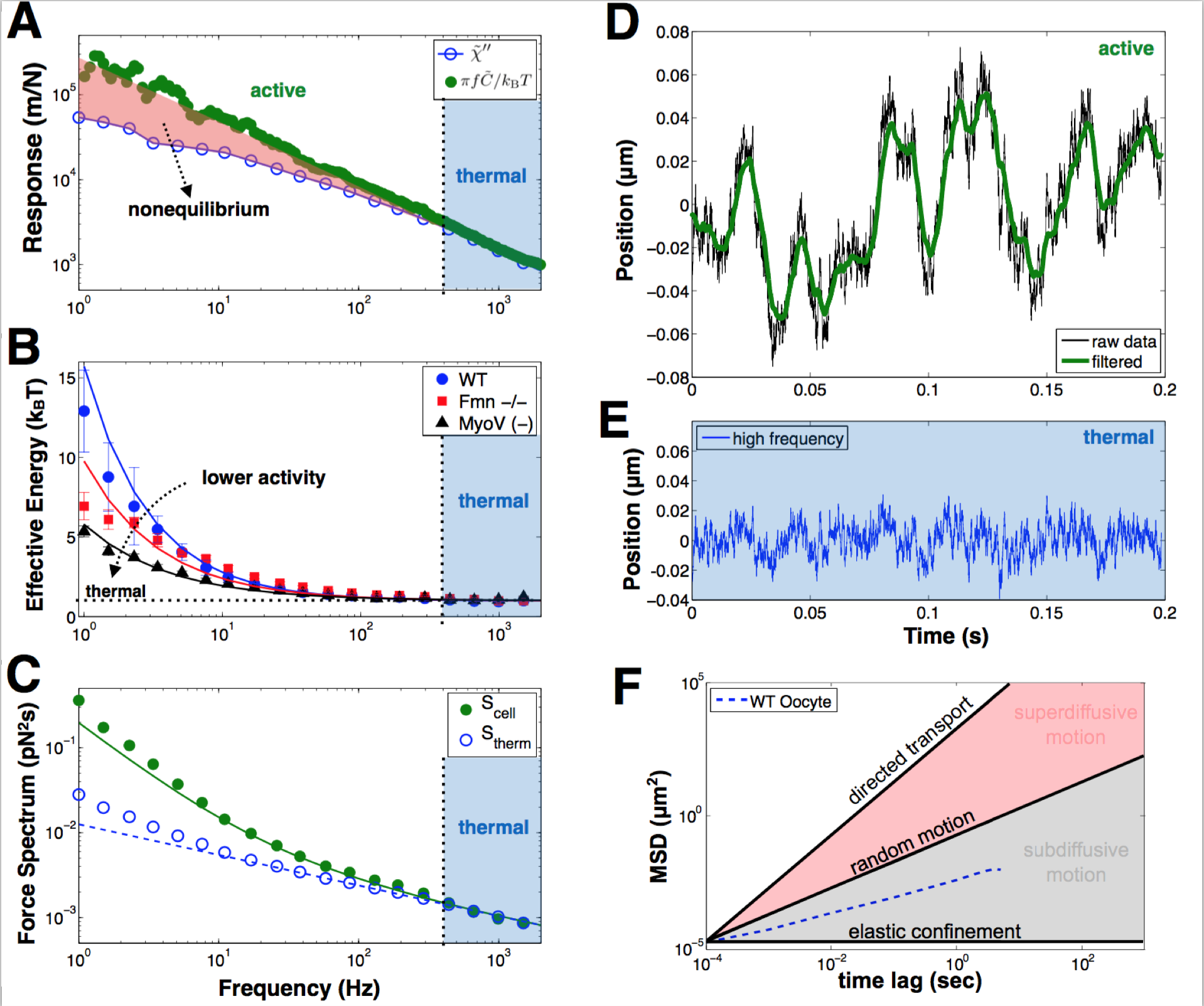}
\caption{{\label{fig:fig3}
\textbf{Active force generation by myosin-V drives the cytoplasmic-skeleton out-of-equilibrium.} (A) At frequencies below 400 Hz the spontaneous motion of vesicles in WT oocytes (green circles) is larger than expected for thermal equilibrium (blue circles). This shows that active forces are contributing to vesicle motion in this regime as highlighted by the red shaded region.
This is direct evidence of nonequilibrium behavior in the cytoplasmic-skeleton
(via violation of the fluctuation-dissipation theorem (FDT)). At high
frequencies the observed vesicle motion resembles thermal motion indicated
by the blue shaded region. (B) WT oocytes (blue) are the furthest
from equilibrium as shown by their higher effective energy. In the
absence of actin (red) or when myosin-V is inactivated (black) the
dynamic actin-myosin-V meshwork is compromised and oocytes have lower effective
energy. Solid lines are theoretical fits (equations in SI, error bars
= SEM). (C) The cell force spectrum ($S_{\mathrm{cell}}$) experienced by
vesicles is the sum of thermal forces ($S_{\mathrm{therm}}$)
and active forces ($S_{\mathrm{active}}$). At high frequencies $S_{\mathrm{cell}}$ (green) is dominated
by $S_{\mathrm{therm}}$ (blue) and the two spectra coincide (blue
shaded region). At lower frequencies $S_{\mathrm{cell}}$ is larger
than $S_{\mathrm{therm}}$ showing the existence of additional active
forces. Solid and dashed lines are theoretical predictions (equations in SI), low-frequency deviation is due to simple power-law model. Standard error of experimental data is within marker size. (D) When a representative trajectory (black) is filtered to remove
the high-frequency thermal fluctuations the result is a smoothed trajectory
(green) that represents actively driven motion. The difference between
the true trajectory (gray) and the smoothed trajectory (green) recovers
the high-frequency thermal fluctuations (blue) shown in (E). (F) The mean-squared-displacement (MSD) of vesicles, calculated from trajectory data, indicates they undergo random-confined motion in the oocyte cytoplasmic-skeleton at short timescales.  This behavior transitions to active diffusion at longer times \cite{25774831}, and is reminiscent of cytoplasmic stirring \cite{24876498}.%
}}
\end{center}
\end{figure}

 We quantify and explain both the passive
    and active contributions driving intracellular fluctuations, by independently measuring the response function and power spectral density.  We use AMR to
    measure $\tilde{\chi}$ and laser-tracking to measure $\tilde{C}$,
    and use this information to check for violation of FDT \cite{17234946,11724945,Fodor_2015},
    which indicates active force generation. In oocytes, at high-frequencies
    the mechanical response ($\tilde{\chi}^{\prime\prime}$) and the spontaneous vesicle
    motion ($\pi f\tilde{C}/k_{B}T$) are related by the FDT as expected for thermal
    fluctuations \cite{24876498,25126787} (Fig. \ref{fig:fig3}A). At frequencies below $\sim400$ Hz, the observed motion of vesicles
    is dominated by an active energy consuming process (highlighted
    by the pink shaded region between the two curves in Fig \ref{fig:fig3}A).  
    
     To quantify the nonequilibrium activity in an active soft material
    it is instructive to consider the effective energy \cite{11724945,19717437,21770546,19792774}, $E_{\mathrm{eff}}=\pi f\tilde{C}/\tilde{\chi}^{\prime\prime}$,
    which is a measure of how far the system is from thermal equilibrium.  Note that the measurement of  $E_{\mathrm{eff}}$ requires two separate measurements: rheology to get $\tilde{\chi}^{\prime\prime}$ and laser interferometry to get $\tilde{C}$.
    WT oocytes exhibit a strong departure from equilibrium due mainly to the actin-myosin-V network activity. Accordingly, the
    deviation is reduced when either actin is absent (Fmn-/-) or myosin-V
    is inactivated (MyoV (-)) (Fig \ref{fig:fig3}B). This
    is quantitative confirmation that the dynamic actin-myosin-V meshwork
    drives vesicle dynamics out-of-equilibrium in the cytoplasmic-skeleton
    of mouse oocytes \cite{23873150,25774831}. Note that in Fmn-/- and MyoV(-) not all activity is abolished,  this is presumably due to residual actin-myosin-V that is active or other remaining sources of non-thermal activity in the oocyte (e.g., enzymes, polymerization, myosin-II in the cortex \cite{23873150}.
    
    To develop a more intuitive picture of the activity we quantify the
    forces generated in the cell by calculating the cell force spectrum
    ($S_{\mathrm{cell}}$).
    $S_{\mathrm{cell}}$ directly represents the average total force
    on a vesicle from all stochastic sources (active and thermal) inside the cell, and has been used by several previous studies to quantify activity \cite{14611619,25126787,Gallet_2009,Fodor_2015}.  The force spectrum is a convenient and common way to quantify the total intracellular forces.
    In analogy to the force on a simple spring, where the force is the
    stiffness multiplied by displacement ($F=\kappa\Delta x$, where $\kappa$
    is stiffness and $\Delta x$ is displacement), we calculate $S_{\mathrm{cell}}=(6\pi R)^{2}|G^{*}|^{2}\tilde{C}$
    where $\vert G^{*}\vert^{2}$ represents the stiffness of the cytoplasmic-skeleton \cite{25126787,Gallet_2009,25375540}.
    In our framework we separate the total force spectrum in the cell to be the sum of thermal forces and active forces ($S_{\mathrm{cell}}=S_{\mathrm{therm}}+S_{\mathrm{active}}$) \cite{19658739,Fodor_2015}. The thermal force spectrum is calculated directly from the AMR measurements as $S_{\mathrm{therm}}=12\pi RG^{\prime\prime}k_{B}T/\omega$, where $R$ is the vesicle radius, $G^{\prime\prime}$ is the dissipative component of the shear modulus, and $\omega=2\pi f$ is frequency. $S_{\mathrm{therm}}$  represents the forces present at thermal equilibrium and is a direct consequence of the FDT \cite{Kubo_1966}.  The resulting force spectrum is shown in Fig \ref{fig:fig3}C where at high frequencies the cell force spectrum ($S_{\mathrm{cell}}$) is dominated by thermal forces (where $S_{\mathrm{active}} \ll S_{\mathrm{thermal}}$) as expected, but at lower frequencies ($f<400$ Hz) active forces dominate ($S_{\mathrm{active}} \gg S_{\mathrm{thermal}}$) which result in higher total force experienced by vesicles in the cytoplasm over thermal equilibrium (lines represent theoretical model \cite{Fodor_2016} introduced in next section).  To visualize the effect of thermal and active forces we show a representative trajectory of a vesicle (Fig. \ref{fig:fig3}D, black). Based on our measured violation of FDT (Fig. \ref{fig:fig3}A), we apply a filter to isolate the low-frequency ($< 400$ Hz) vesicle motion which is mainly due to active processes, and the high-frequency motion that is dominated by thermal fluctuations (Fig. \ref{fig:fig3}D green, E respectively).

\subsection{Modeling actively driven vesicle motion}

Mouse oocytes in prophase-I exhibit non-directed motion of vesicles in a viscoelastic environment (Fig. \ref{fig:fig3}F, blue) due to a unique cytoplasmic network (Fig. \ref{fig:fig4}A). We use this picture to motivate our theoretical model of generic active forces in the cell cytoplasm and do not impose any physical assumptions on the origin of that force. Therefore, our model lumps all active forces together as a single dominant process, regardless of their origin.
To gain access to the molecular-scale kinetics of the dominant active process
driving the cytoplasmic interior of oocytes we developed a quantitative
model describing vesicles embedded in a viscoelastic environment that
are subjected to thermal and active forces. Here, the nonequilibrium processes provide the active forces that reorganize the polymer cytoskeleton and drive motion of the vesicles. Our theoretical framework
extends previous approaches for near-elastic networks \cite{18232824,25375540,Fodor_2015} to include the complex
mechanical properties measured experimentally (Fig. \ref{fig:fig2}). 

We begin by using the traditional Langevin equation to describe a vesicle freely fluctuating in a viscoelastic continuum where the left side of equation \ref{eq:motion1} is the drag force on the vesicle and the right side includes the elastic caging force from the cytoskeleton (with stiffness $\kappa$) and the thermal force.  It is an application of Newton's 2nd law on the fluctuating vesicle.  Equation \ref{eq:motion1} describes how a vesicle thermally fluctuates randomly about its equilibrium position, but on average will not move from its original position due to elastic caging of the cytoskeleton,

\begin{equation}
\int^{t}\mbox{d}t^{\prime}\gamma(t-t^{\prime}) \dot x (t')=-\kappa(x-x_{0})+\mathbf{\xi}\label{eq:motion1}
\end{equation}
where $x$ is the position of the vesicle, $x_{0}$ is the position of the elastic cage, $\xi$ is thermal force taking the form of Gaussian colored noise with correlations $\langle\xi(t)\xi(t^{\prime})\rangle=k_{B}T\gamma(|t-t^{\prime}|)$, as provided by the FDT \cite{Kubo_1966}, $k_{B}$ is the Boltzmann constant, and $T$ is the bath temperature of the environment. The viscoelastic material properties are modeled as a soft glassy material, where $\kappa$ is the harmonic trap stiffness centered at $x_{0}$, and the memory kernel, is $\gamma(t)=\kappa(\zeta_{\alpha} / t )^{\alpha}/\Gamma(1-\alpha)$ where $\Gamma$ is the Gamma function, $\alpha$ is the viscoelastic power-law frequency dependence,and $\zeta_{\alpha}$ is a passive timescale of the material (see Table 1 for symbol definitions and SM for details).  In an attempt to capture the active vesicle dynamics, we introduce equation \ref{eq:motion2}, which phenomenologically describes how nonthermal forces remodel the cytoplasmic-skeleton and determine the dynamics of the elastic caging,

\begin{equation}
\int^{t}\mbox{d}t^{\prime}\gamma(t-t^{\prime}) \dot x_{0} (t')=\kappa \zeta_{\alpha} v_{\mathrm{A}}\label{eq:motion2}
\end{equation}
where the left side is the drag force on the cage and the right side is the active nonthermal force that drives motion of the cage. Equation \ref{eq:motion2} is an application of Newton's 2nd law on the elastic cage, which is a geometric constraint, and fundamentally does not have mass.  In this case, the cage represents the confining effects of the cytoplasmic skeleton and thus is endowed with the mass of this surrounding material and its center of mass corresponds to the center of the cage. The equilibrium position of the cage is $x_{0}$ and the active force is $\kappa \zeta_{\alpha} v_{\mathrm{A}}$ where $v_{\mathrm{A}}$ is the velocity of the cage motion.  For completeness, equation \ref{eq:motion2} should include the force applied on the cage by the vesicle (the action-reaction principle) and the thermal force on the cage itself.  However, when each increment of vesicle motion is small compared to the cage size, these forces do not affect the cage dynamics to leading order and previous work has suggested they can be neglected (see equation 3a,b in \cite{25375540}). The overall dynamics describe a vesicle in the cytoplasmic-skeleton that experiences an elastic caging  (equation \ref{eq:motion1}), until active forces act to remodel the cytoskeleton and move the cage (equation \ref{eq:motion2}), resulting in a vesicle motion driven by both thermal and active forces. This can be considered a coarse-grained approach where the dynamics are divided into two equations of motion describing the thermal fluctuations (equation 1) and the active motion (equation 2) separately to simplify the mathematics and subsequent analysis.   Therefore, this phenomenological model can describe diffusive-like motion driven by nonthermal forces in a viscoelastic medium, also known as active diffusion.  Interestingly, a recent study used a coarse-grained two-fluid continuum model \cite{Yasuda_2017} and arrived at the same conclusion \cite{Fodor_2015}. 

The parameters characterizing the mechanical properties are obtained from AMR measurements. The memory kernel quantifies the mechanical resistance of the cytoplasmic-skeleton and influences the dynamics of both the vesicle, $x$, and the cage, $x_{0}$.  In principle, one would expect the memory kernel in equations \ref{eq:motion1} and \ref{eq:motion2} to be different since equation \ref{eq:motion1} describes motion of the vesicle, and equation \ref{eq:motion2} describes motion of the cage.  Provided that the dissipation is mediated by the surrounding medium in both cases, then the memory kernels would only differ by a prefactor. In this model the prefactor ends up on both sides of equation \ref{eq:motion2} and mathematically cancel, leaving the memory kernels in equations \ref{eq:motion1} and \ref{eq:motion2} identical (see previous derivation \cite{25375540}).

The microscopic interpretation of $\kappa$ is the local caging due to the cytoskeleton (as evident
from the low frequency plateau in the measured shear modulus Fig. \ref{fig:fig2}), where $x_{0}$ is the location of this cage. In the
absence of active forces, the position of this local cage does not
move ($\dot x_{0}=0$). When active forces do exist, as in
living cells, they rearrange the cytoskeleton and drive motion of
the cage ($\dot x_{0} \neq0$). The active force experienced by the cage results from the active burst velocity, $v_{A}$,
which is a stochastic process consisting of alternating active and
quiescent periods. During an active phase the active burst velocity,
$v_{A}$, takes on a constant non-zero value, $v$, for a random exponentially distributed time of average value $\tau$.  During a quiescent phase the active burst velocity is 0 for an exponentially distributed random time of average value $\tau_{0}$ \cite{25375540,Fodor_2015,21770546}. A schematic example is shown in Fig. 4B, right inset. The active burst velocity is a zero-mean stochastic process with correlations $\langle v_{A}(t)v_{A}(0)\rangle=k_{B}T_{A}\exp(-|t|/\tau)/(\kappa\zeta_{\alpha}\tau)$, where $k_{B}T_{A}=\kappa\zeta_{\alpha}(v\tau)^{2}/[3(\tau+\tau_{0})]$ defines an active energy scale. Note that the dynamics of the joint process $\{x,x_{0}\}$ is not Markovian because of the memory kernel in the equations of motion.

From the generalized Stokes-Einstein relation, the memory kernel can be associated with a complex shear modulus $G^{*}=G^{\prime}+iG^{\prime\prime}$ of the form
\begin{equation}
G^{\prime}(\omega)=\frac{\kappa}{6\pi R}\left[ (\omega\zeta_{\alpha})^{\alpha}\cos\left(\frac{\pi\alpha}{2}\right)+1\right]\label{eq:G'}
\end{equation}
\begin{equation}
G^{\prime\prime}(\omega)=\frac{\kappa}{6\pi R} (\omega\zeta_{\alpha})^{\alpha}\sin\left(\frac{\pi\alpha}{2}\right)\label{eq:G''} .
\end{equation} 
This model separates the passive forces that originate from the thermal fluctuations of the medium and the active forces that depend on the the energy consuming processes in the active material.  Note that if the active force on the cage is zero, ($\kappa \zeta_{\alpha} v_{\mathrm{A}} = 0$), then equation \ref{eq:motion2} is zero, and the equation of motion simplifies to confined viscoelastic Brownian motion (equation \ref{eq:motion1}).  If the active force is not zero, then it originates from an active process that has step-like velocity kinetics (Fig \ref{fig:fig4}B, right inset). These force kicks drive the nonequilibrium fluctuations in our model.

The nonequilibrium properties of the active force are quantified
by the deviation from the FDT defined
by a frequency dependent effective energy, $E_{\mathrm{eff}}$, as discussed earlier. We compute it in terms of the microscopic
ingredients of our theoretical model,
\begin{equation}
E_{\mathrm{eff}}(\omega)=k_{B}T+\frac{1}{(\omega\zeta_{\alpha})^{3\alpha-1}\sin(\pi\alpha/2)}\frac{k_{B}T_{A}}{1+(\omega\tau)^{2}}\label{eq:Eeff}
\end{equation}
where $k_{B}T_{A}$ is the energy scale associated with the active
process. The dynamics of the tracer can be written as:
$x(t)=\int^{t}\mbox{d}t^{\prime}\chi(t-t^{\prime}) F_{\mathrm{cell}}(t^{\prime})$,  
where $F_{\mathrm{cell}}=\xi+\kappa x_{0}$ is the cell force which describes the thermal forces arising in the cell and the effect of the active forces via $x_{0}$. Explicity, we calculate the cell force spectrum as $S_{\mathrm{cell}}=\langle\vert\tilde{F}_{\mathrm{cell}}\vert^{2}\rangle=\langle\vert\tilde{\xi}\vert^{2}\rangle+\kappa^{2}\langle\vert\tilde{x}_{0}\vert^{2}\rangle$
which corresponds to spectrum calculated previously $S_{\mathrm{cell}}=\tilde{C}/\vert\tilde{\chi}\vert^{2}=(6\pi R)^{2}\vert G^{*}\vert^{2}\tilde{C}$ from experiments.
Since we separate the thermal and active contributions to the force spectrum in our model, we compute the explicit expressions for each,
\begin{equation}
S_{\mathrm{therm}} (\omega) =2 k_{B}T \kappa \zeta_{\alpha} (\omega\zeta_{\alpha})^{\alpha-1}\sin\left(\frac{\pi\alpha}{2}\right)\label{eq:S_T}
\end{equation}
\begin{equation}
S_{\mathrm{active}} (\omega) =\frac{2\kappa\zeta_{\alpha}}{(\omega\zeta_{\alpha})^{2\alpha}}\frac{k_{B}T_{A}}{1+(\omega\tau)^{2}}\label{eq:S_A}
\end{equation}
The active force spectrum (equation \ref{eq:S_A}) can now be used to extract the kinetics of the active force.  Details of the derivation can be found in the SM.

%
%
%
%


\begin{center}
Table 1. List of symbols, variables and definitions

\begin{tabular}{|c|c||c|c||c|c|}
\hline 
$\tilde{\chi}(\omega)$ [m/N] & response function & $T$ [K] & bath temperature \tabularnewline
\hline
$\omega$ [rad/s] & frequency & $T_A$ [K] & active temperature \tabularnewline 
\hline
$\gamma(t) [N/m] $ & memory kernel & $\tau$ [s] & active persistence time \tabularnewline
\hline
$G^*(\omega)$ [N/m$^2$] & complex shear modulus & $\tau_0$ [s] & active waiting time \tabularnewline
\hline
$G'(\omega)$ [N/m$^2$] & elastic modulus & $v$ [m/s] & active burst amplitude \tabularnewline
\hline
$G''(\omega)$ [N/m$^2$] & viscous modulus & $\tilde{C}(\omega)$ [m$^2$-s] & position power spectral density \tabularnewline
\hline
$R$ [m] & vesicle radius & $E_\text{eff}(\omega)$ [N-m] & effective energy   \tabularnewline
\hline
$\kappa$ [N/m] & local cage stiffness & $\tilde{F}_\text{cell}(\omega)$ [N-s]& total cell force \tabularnewline
\hline
$\zeta_{\alpha}$ [s] & passive timescale & $S_\text{cell}(\omega)$ [N$^2$-s] & cell force spectrum  \tabularnewline
\hline 
$\alpha$ & power-law exponent & $S_\text{thermal}(\omega)$ [N$^2$-s] & thermal force spectrum  \tabularnewline
\hline 
$x(t)$ [m] & vesicle position & $S_\text{active}(\omega)$ [N$^2$-s] & active force spectrum \tabularnewline
\hline
$x_0(t)$ [m] & local cage position & $F$ [N] & active force amplitude \tabularnewline
\hline
$\xi(t)$ [N] & thermal noise & $v_\text{vesicle}$ [m/s] & average vesicle velocity \tabularnewline
\hline
$v_A (t)$ [m/s] & active burst noise & $\Delta x$ [m] & average step-displacement \tabularnewline
\hline

\end{tabular}

\label{tabular:Theoretical-symbols}

\end{center}

\subsection{Extracting \textit{in-vivo} molecular-scale force kinetics}

\begin{figure}[H]
\begin{center}
\includegraphics[width=0.7\columnwidth]{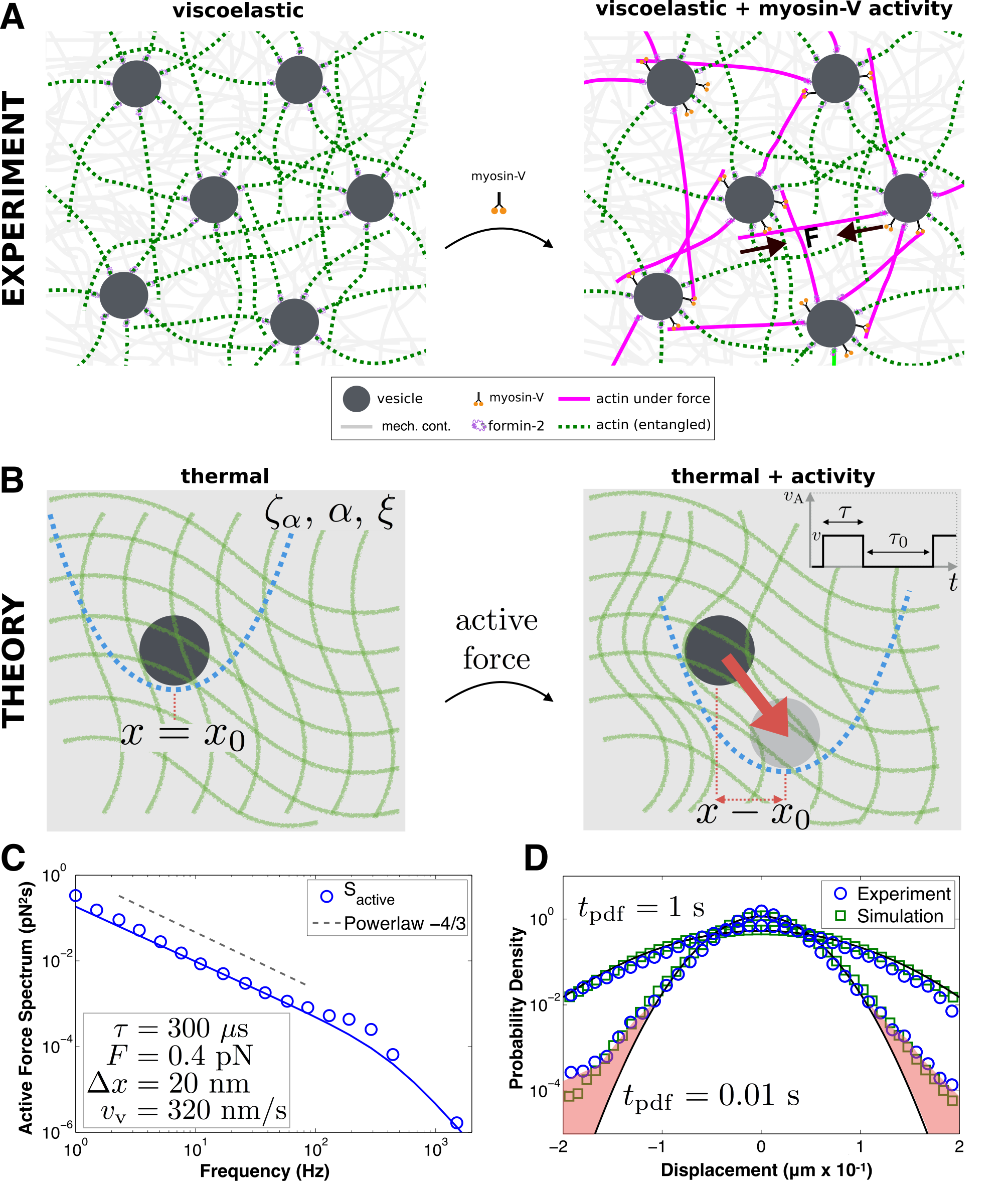}
\caption{{\label{fig:fig4}
\textbf{A theoretical model of active mechanics connects \textit{in vivo} measurements
to molecular force kinetics.} (A) Vesicles (dark gray) are embedded in the complex mechanical continuum (gray background) of the oocyte interior. Actin filaments emanate from the the surface of vesicles creating an entangled network (green, left). Myosin-V motors generate force on actin (magenta, right) giving rise to forces throughout the network driving random motion of vesicles.  (B) We model vesicles embedded in a mechanical continuum
with local cage stiffness ($\kappa$) represented by the blue harmonic potential, viscoelastic dissipation ($\zeta_{\alpha},\alpha$), and thermal fluctuations $(\xi$) (left). Active processes rearrange the network through bursts of motion ($v_{A}$), resulting in displacement of the local cage (blue harmonic), which generates the active force ($\kappa \zeta_{\alpha} v$, red arrow) that drives vesicle motion towards the local minimum (right)(inset indicates active force kinetics). (C) The active force spectrum ($S_{\mathrm{active}}$)
quantifies the forces on vesicles due to only active processes. Combined
with our quantitative model we find that the vesicles are subject
to 0.4 pN of force, during a power-stroke of length $\Delta x \sim 20$ nm and duration 300 $\mu$s, resulting
in a vesicle velocity of 320 nm/s, which is strikingly similar to
the kinetics measured for single molecule myosin-V \textit{in-vitro} and the \textit{in-vivo} vesicle velocity. The solid line is the theoretical fit (equation \ref{eq:S_A}), and the dotted line is a $-4/3$ power-law consistent with the cytoplasmic-skeleton
mechanics. (D) Simulated vesicle motion (green squares) agrees with experimental data (blue circles) for a wide range of timescales as shown by the probability distribution of displacements. This includes long timescale ($t_{\mathrm{pdf}}=1$ s) Gaussian behavior and short timescale ($t_{\mathrm{pdf}}=0.01$ s) non-Gaussian tails (indicated by red-shaded regions) that suggest molecular motor behavior, where $t_{\mathrm{pdf}}$ is the time-lag
for calculation of the displacement correlations. Gaussian distributions
shown in black. These results show that molecular level kinetics of
active processes can be extracted from mesoscopic \textit{in-vivo} measurements.%
}}
\end{center}
\end{figure}

By combining our analytical model and experimental measurements, we extract
the force kinetics driving active mechanical processes in
the oocyte. First, the mechanical measurements from AMR were fit to equations \ref{eq:G'} and \ref{eq:G''} to determine the mechanical parameters: $\alpha,\zeta_{\alpha},\kappa$ from the data shown in Figure \ref{fig:fig2}A, B. Once the mechanics is determined, the remaining equations are largely constrained. The best fit of the data for effective energy (Figure \ref{fig:fig3}B) is used to determine $k_{B}T_{A}$ in equation \ref{eq:Eeff}, where the fit is independent of $\tau$ when $\tau<10$ ms. From fitting the active force spectrum we find that $\tau=0.3$ ms (the only free parameter) to capture the high-frequency drop-off (Figure \ref{fig:fig4}C). The fit parameters for the theoretical model applied to all experimental conditions are shown in Table 2. To summarize: $\alpha,\zeta_{\alpha},\kappa$ are obtained from fitting AMR measurements of the mechanics ($G^{\prime},G^{\prime\prime}$), $T_{A}/T$ is obtained from fitting the effective energy data ($E_{\mbox{eff}}$), and $\tau$ is obtained from fitting the active force spectrum ($S_{\mathrm{active}}$). Our model  extracts the molecular-scale force kinetics directly from fitting the active force spectrum. As a result, we are able to capture the short timescale power-stroke ($\tau$) of the active process, which is not possible from fitting the long timescale plateau of the MSD as done previously \cite{25126787,24876498}

To investigate how well our estimated kinetic parameters describe our experimental data we turn to Brownian dynamics simulations.  We use the measured parameters (mechanics, kinetics, etc.) extracted from the AMR measurements and force spectrum to simulate vesicle dynamics via a stochastic equation of motion (see SM) using previously developed methods \cite{Bochud_2007,Baczewski_2013}. These simulations of vesicle motion use exclusively values measured from experiment, without any free parameters. At this point, it is important to note that the fit parameters in Table 2 are based on first and second moment analysis from experimental data and they do not capture \textit{a priori} the full distribution of vesicle dynamics, especially the non-Gaussian behavior which depends on the details of higher-order moments of the active burst statistics.

To test this, we analyze the simulations by calculating the full probability distribution of the simulated vesicle displacements (also known as van Hove Correlations \cite{Betz_2014}) and compare them to our experimental measurements (Fig \ref{fig:fig4}D). We find that the statistics of the simulated vesicle motion is in excellent agreement with the measured spontaneous motion of vesicles in living oocytes. Measured and simulated motion of vesicles is nearly identical including the central Gaussian region and the non-Gaussian tails (shaded red in Fig \ref{fig:fig4}D) at short timescales. This non-Gaussian behavior suggests molecular motor activity that is dominated by the single most persistent motor closest to the vesicle \cite{Toyota_2011}. While our analytic model describes the mean values of the experimental measurements, these simulations exploit the full range of data by capturing the entire distribution of displacement correlations from experiments.  In other words there was no guarantee simulations would agree with the experiments, particularly for the tails of the distribution (or any other non-Gaussian behavior).  Hence, the agreement found between simulations and experiment is non-trivial and supports the estimated kinetic parameters and use of the memory kernel acting on both the vesicle and the cage. These results show that our model of step-like active forces captures the overall motion of vesicles (including higher-order statistics) in the cytoplasmic-skeleton of living oocytes.

\begin{center}
Table 2. Theoretical fit parameters

\begin{tabular}{|c|c|c|c|c|c|}
\hline 
 & \multicolumn{3}{c|}{mechanical properties} & \multicolumn{2}{c|}{activity}\tabularnewline
\hline 
\hline 
 & $\alpha$ & $\zeta_{\alpha}$ [s] & $\kappa$ [pN$\cdot\mu$m$^{-1}$] & $T_{A}/T$ & $\tau$ [ms]\tabularnewline
\hline 
WT & 0.64 & 0.0657 & 20 & 5.5 & 0.30\tabularnewline
\hline 
Fmn-/- & 0.6 & 0.103 & 18 & 5.0 & 0.15\tabularnewline
\hline 
MyoV (-) & 0.57 & 0.163 & 27 & 3.8 & 0.1\tabularnewline
\hline 
 & \multicolumn{3}{c|}{from AMR } & \multicolumn{2}{c|}{from AMR/PMR}\tabularnewline
\hline

\end{tabular}

\label{tabular:Theoretical-fit-parameters}

\end{center}

\section{Discussion}

\subsection{Myosin-V activity maintains a soft cytoplasmic-skeleton}

Our mechanical measurements (AMR) indicate that when myosin-V activity is present the cytoplasmic-skeleton in oocytes is softer.  This observation was first reported by Almonacid et al \cite{25774831}, and the new data in this study extends this observation across a wide range of frequencies.  Additionally, the combined measurements of viscoelastic properties and spontaneous vesicle motion provides access to an independent measurement of mechanics and activity. The mechanism of how softening occurs is not yet clear but may be related to the structural connectivity of the actin-myosin-V network.  In our experiments, it is possible that when myosin-V is inactivated (MyoV(-)) it serves as a passive cross-linker that stiffens the environment or cross-links the vesicle to the network. Since cross-linking leads to highly nonlinear stiffening \cite{Sharma_2016} (if crosslink distance is not large compared to the persistence length), a small change in cross-linking could effectively increase the stiffness of the cytoplasmic-skeleton.  Alternatively, the presence of myosin motors may reduce the mechanical contribution of the actin cytoskeleton by causing fluidization (see \cite{Oriola_2017}) so that other intracellular structures (e.g. intermediate filaments) dominate the mechanics in the WT condition. This interpretation is also consistent with the finding that the removal of the cytoplasmic actin has no effect on the shear modulus. However, in the absence of experimental tools to dissect the mechanical contribution of all cytoplasmic components in the mouse oocyte, the answer to this question is currently out of reach.

\subsection{Active force spectrum reveals molecular-scale kinetics of the active process}
Myosin-V is typically considered a directed transport motor.  However, in prophase-I mouse oocytes, myosin-V drives vesicle motion in a non-directed fashion as evidenced by its sub-linear scaling of the mean-squared-displacement (MSD) (Fig. \ref{fig:fig3}F, blue).  Previous work has also shown that myosin-V drives random motion of vesicles (active diffusion) on longer timecales of minutes \cite{25774831}. To further illustrate this point, Fig. \ref{fig:fig4}A(left) shows a schematic illustration of vesicles and actin filaments embedded in the oocyte interior. The complex mechanical environment is represented by a generic continuum (gray background). When myosin-V is attached to a vesicle and applies force on an actin filament, this force is transduced through the actin filament to a neighboring vesicle where it is bound by formin-2 (e.g. black arrows in Fig. \ref{fig:fig4}A, right). This process occurs frequently throughout the oocyte interior resulting in randomly distributed force-dipoles (similar to myosin-II in some systems \cite{24876498,17234946}). Thus, each vesicle experiences forces in random directions due to the action of myosin-V throughout the network.  This is represented by many actin filaments under force as shown in magenta in Fig. \ref{fig:fig4}A(right).  Therefore, due to the unique network connectivity in the cytoplasmic-skeleton, myosin-V motors drive active random motion in prophase-I mouse oocytes reminiscent of cytoplasmic stirring observed previously that was driven by myosin-II \cite{24876498}.  This actin-myosin-V activity is expected to be the dominant driver of active diffusion in our system \cite{25774831}.  

By looking at the active force spectrum in WT oocytes, we can characterize the kinetics of the dominant active process. The first factor in equation \ref{eq:S_A} depends on the mechanical properties of the cytoplasmic-skeleton while the second factor is related to the kinetic properties of the active processes.  This illustrates that the active force spectrum is dependent on the active force generation as well as the mechanics of the environment it must push against. The power-law scaling of the active force spectrum contains information about the underlying physics.  At lower frequencies the active force spectrum scales as $f^{-2\alpha}$ reflecting the active forces pushing against the viscoelastic environment. At higher frequencies the active force spectrum is dominated by the kinetics of the active force generation process which scales as $f^{-(2\alpha + 2)}$. This is illustrated in Fig. S3 where at low frequencies there is a power-law dependence of -4/3 consistent with the mechanics ($\alpha\approx2/3$) and at high frequencies the power-law dependence is approximately -10/3 consistent with kinetics and mechanics. The crossover between these two regimes occurs at $1/\tau$ as shown approximately by the purple dotted line (Fig. S3).

Our predicted active force spectrum exhibits power-law scaling that is dependent on the mechanical properties of the system ($\alpha$), in contrast to previous studies \cite{18232824,14611619,Levine_2009}. Previous theoretical developments predict an active force spectrum that scales as $f^{-2}$, which is independent of the mechanical properties \cite{14611619}, and a plateau ($f^{0}$) below a critical frequency (equation 2 in \cite{24876498} and equation S1 in \cite{25126787}). It is worth noting that in the near-elastic case ($\alpha \sim 0$), our model recovers the behavior observed in previous work \cite{25126787,24876498}. The divergence of the active force spectrum at low frequencies is consistent with our experimental measurements, and others, where a low-frequency plateau is not observed \cite{18764230,Gallet_2009,Robert_2010,25126787}.

The experimental measurements for the active force spectrum ($S_{\mathrm{active}}=S_{\mathrm{cell}}-S_{\mathrm{therm}})$
and theoretical model are compared in Fig \ref{fig:fig4}C.
The low frequency power-law behavior is clearly seen, while at higher
frequencies the active forces drop off rapidly which may reflect
molecular motor statistics \cite{10944217}. This allows us to calculate other kinetic parameters, such as the apparent active force amplitude, $F$, which is the typical force amplitude that an endogenous vesicle feels that drives its non-thermal motion. 

\begin{equation}
F = \left[3\kappa k_{B}T_{A}(\tau+\tau_{0})/\zeta_{\alpha}\right]^{1/2} \label{eq:F_active}
\end{equation}
where $\tau_{0}$ is deduced from the single-molecule myosin-V duty cycle \cite{17878301} (details in SM).  Here, $F$, lumps all activity into one value to quantify the active force. Once this force is known it is also straight-forward to calculate the typical vesicle velocity,

\begin{equation}
v_{\mathrm{vesicle}}=\frac{F}{\kappa\zeta_{\alpha}} \label{eq:vel_ves}
\end{equation}
which is a ratio of the driving force, and the resistance provided by the surrounding environment, $\kappa \zeta_{\alpha}$. Following this argument the typical vesicle displacement induced by one force kick can be computed as $\Delta x=F /\kappa$.  Thus, once the timescale of the force kick ($\tau$) is extracted from equation \ref{eq:S_A}, the apparent force felt by the vesicle ($F$), the expected vesicle velocity ($v_{\mathrm{v}}$), and the step-size $\Delta x$ can be deduced.

Our combined experimental and theoretical framework allows access to the molecular-scale kinetics via mesoscopic measurements of the active force spectrum (Fig. \ref{fig:fig4}C). The active force spectrum, $S_{\mathrm{active}}$, has only one free fitting parameter, $\tau$, which corresponds to the molecular-scale activity.  All other model parameters are fixed previously. In living oocytes we find that endogenous vesicles experience a force of $F \sim0.4$ pN during a power-stroke of $\tau \sim300$ $\mu$s duration, with a step-size of $\Delta x \sim 20$ nm. This is strikingly similar to single molecule myosin-V kinetics measured \textit{in-vitro} (see Table 3)\cite{10448864,17878301,10944217,16100513,15286720,20418880,11740494}. In addition, we find that the predicted average vesicle velocity due to active forces is $v_{\mathrm{vesicle}} \sim320$ nm/s which is in agreement with myosin-V velocity \textit{in vitro} \cite{10448864,16100513,15286720}, as well as the velocity of myosin-V driven vesicles measured in \textit{in-vivo} oocytes via video microscopy \cite{25774831,21983562}. While our model does not assume the action of a single motor acting on the vesicle, the extracted kinetics are in agreement with single myosin-V. Since it is likely there are several motors affecting the vesicle dynamics, our results suggest that the myosin-V motors in prophase-I oocytes do not act simultaneously, thus allowing individual motor kicks to be observed.  Interestingly, previous work has estimated the number of motors involved based on quantification of energy dissipation \cite{Fodor_2016}.  It is important to note that while our extracted kinetics are in agreement with myosin-V measurements, this is not sufficient to prove that we are accessing single-molecule kinetics.  Together, our results suggest that force kinetics of \textit{in-vitro} myosin-V is remarkably similar to the active force fluctuations in \textit{in-vivo} oocytes, where they drive the composite cytoplasmic-skeleton out-of-equilibrium (Table 3).

\begin{center}

%

Table 3. Molecular-scale force kinetics

\begin{tabular}{|c|c|c|c|}
\hline 
 & \textit{in-vivo} oocytes & single-molecule myosin-V & reference\tabularnewline
\hline 
\hline 
$F$ [pN] & 0.4 & 0-4 & \cite{10944217,17878301,10448864,16100513}\tabularnewline
\hline 
$\tau$ [$\mu$s] & 300 & 160-1000 & \cite{17878301,20418880,15286720} \tabularnewline
\hline 
$\Delta x$ [nm] & 20 & 15-25 & \cite{17878301,11740494,20418880,15286720}\tabularnewline
\hline 
$v_{\mathrm{vesicle}}$ [nm/s] & 320 & 270-480 & \cite{10448864, 16100513, 15286720}\tabularnewline
\hline

\end{tabular}
\label{tabular:tab-kinetics}

\end{center}

\subsection{Success and limitations of the current theory and experiments}
Our phenomenological model is designed to capture the nonequilibrium fluctuations of vesicles in the oocyte cytoplasm.  Specifically, this model is able to reproduce the deviation from equilibrium as measured by the effective energy and active force spectrum.  In that respect, the agreement between model and experiments supports the underlying phenomenological picture that: (i) active fluctuations of the vesicles are mediated by the surrounding medium; (ii) the complexity of the surrounding medium can be reduced to a coarse-grained representation described by local caging, (iii) the damping force acting on the vesicle and the cage exhibit similar memory effects.  While this minimal approach is completely phenomenological, several studies have shown that introducing a fictitious cage coordinate has been a powerful framework for describing intermittent dynamics of a particle in a damped medium \cite{Lasanta_2015}. And in the case of mouse oocytes, it is able to capture the nonequilibrium fluctuations of vesicles and predict force kinetics that are in agreement with single-molecule experiments on myosin-V motors.  Nevertheless, this coarse-grained approach inherently neglects the microscopic details of the underlying processes.

The main drawbacks of our minimal model are: (i) We neglect any collective effects that may be occurring due to multiple motor interactions.  If several motors are interacting to generate a force kick, we only capture their combined action and no information about individual motors. (ii) All material properties are taken as an input to the model (as measured from experiments).  Thus, the model is not able to capture how the motors may affect the material properties (e.g. cannot explain softening/stiffening \cite{Ahmed_pnas}).  Therefore it is critical that the material properties of the medium are measured \textit{in-situ}, and the quality of the measurements will strongly affect the model. (iii) Our coarse-grained interpretation involves representing the cytoskeleton as a local cage constraining vesicle motion.  Active motion of the local cage obeys Newton's second law, which typically describes the motion of a specific material point and its associated mass.  In this case, since the local cage is not an actual physical object, it's mass is associated with the material that is surrounding the vesicle.  This interpretation is necessary to apply Newton's second law to the active motion in our phenomenological model, and similar approaches  inspired by the itinerant oscillator model have been used previously \cite{Lasanta_2015}. Together, these drawbacks mask the underlying microscopic processes and future work will focus on gaining access to these details.

To spur future developments and improve this type of analysis we would like to point out the current limitations of the theory and experiments. \textit{In theory:} our phenomenological approach to describe the cage dynamics (equation 2) does not track a specific material point in the cytoskeleton, making it difficult to interpret the physical basis of the memory kernel; the active process driving nonequilibrium motion is dominated by a single timescale, which may not be true in other cell types; and the number of motors driving the vesicle dynamics is not specified, since this information is typically not known \textit{in-vivo}.  Overcoming these limitations in the theory is challenging, since tracking a specific material point in an active material (that is constantly changing) may not be possible, and introducing multiple timescales and details about motors requires more specific information about the experimental system at hand.  A useful first step could be to develop a new metric to characterize the mechanical response of active materials that does not rely on the material itself remaining invariant. \textit{In experiments:} force calibration relies on the high-frequency collapse of fluctuation and response, which requires precise measurement that is often not accessible in many systems; measurements of vesicle motion are limited to a relatively small range ($\sim 400$ nm), due to the linear range of the QPD; and rheological measurements are limited to frequencies greater than $\sim 1$ Hz, due to large-scale motion in the cell.  These experimental limitations could be overcome by implementing new force calibration techniques and a feedback system which continually re-centers the optical trap on the vesicle.

\section{Conclusion}
We quantify the molecular-scale force kinetics of active diffusion in mouse oocytes via experiments, theory, and simulation.  We find that active forces in \textit{in-vivo} oocytes are remarkably similar to myosin-V kinetics \textit{in-vitro}, and that the \textit{in-vivo} kinetics can be extracted from cytoplasmic fluctuations. Our results demonstrate a framework for connecting cellular scale phenomena to their underlying molecular force kinetics \textit{in-vivo} and provide insight on the kinetic origin of active diffusion in mouse oocytes. 

\section{Author contributions}
WWA and TB conceived and supervised the project. WWA, MB, MA, and MHV performed experiments. \'{E}F, NSG, PV, and FvW developed the theoretical model and performed simulations.  WWA and \'{E}F integrated experiment and theory. All authors contributed to data analysis and/or interpretation.  WWA and TB wrote the manuscript, which was seen and corrected by all authors.

\section{Acknowledgements}
We thank Jacques Prost, C\'{e}cile Sykes, Julie Plastino, and Jean-Francois Joanny for helpful discussions. We thank Melina Schuh (MRC Cambridge) for providing the MyoVb tail plasmid, Cl\'{e}ment Campillo for help with synthetic vesicle experiments, and Amanda Remorino for critical reading of the manuscript. WWA is a recipient of post-doctoral fellowships from La Fondation Pierre-Gilles de Gennes and Marie Curie Actions. MA is a recipient of post-doctoral fellowships from the Ligue Nationale contre le Cancer and from the Labex MemoLife. MB is a recipient of an AXA Ph.D. fellowship. NSG gratefully acknowledges funding from the ISF (grant no. 580/12). MHV gratefully acknowledges the Ligue Nationale Contre le Cancer (EL/2012/LNCC/MHV). TB was supported by the French Agence Nationale de la Recherche (ANR) Grants ANR-11-JSV5-0002, and the Deutsche Forschungsgemeinschaft (DFG), Cells-in-Motion Cluster of Excellence (EXC 1003 -- CiM), University of M\selectlanguage{ngerman}ü\selectlanguage{english}nster, Germany.

\bibliographystyle{unsrt}
\bibliography{converted_to_latex.bib%
}

\end{document}